# Storing High-Dimensional Quantum States in a Cold Atomic Ensemble

Dong-Sheng Ding, Wei Zhang, Zhi-Yuan Zhou, Jian-Song Pan, Guo-Yong Xiang, Bao-Sen Shi,[*] and Guang-Can Guo

*Key Laboratory of Quantum Information, University of Science and Technology of China, Hefei 230026, China*

[*]*Corresponding author: drshi@ustc.edu.cn*

The reversible transfer of the quantum information between a photon, an information carrier, and a quantum memory with high fidelity and reliability is the prerequisite for realizing a long-distance quantum communication and a quantum network [1, 2]. Encoding photons into higher-dimensional states could significantly increase their information-carrying capability and network capacity [3, 4]. Moreover, the large-alphabet quantum key distribution affords a more secure flux of information [5]. Quantum memories have been realized in different physical systems, such as atomic ensembles [6-11] and solid systems [12-15] etc., but to date, all quantum memories only realize the storage and retrieval of the photons lived in a two-dimensional space spanned for example by orthogonal polarizations [6-15], therefore only a quantum bit could be stored there. Here, we report on the first experimental realization of a quantum memory storing a photon lived in a three-dimensional space spanned by orbital angular momentums via electromagnetically induced transparency in a cold atomic ensemble. We reconstruct the storage process density matrix with the fidelity of 85.3% by the aid of a 4-f imaging system experimentally. The ability to store a high-dimensional quantum state with high fidelity is very promising for building a high-dimensional quantum network.

A main goal of quantum communication is the development of a quantum network through which users can exchange quantum information at will. Such a network would consist of spatially separated quantum memories used to store and manipulate information, and quantum channels through which different quantum memories could be connected. Photons are robust and efficient carriers of quantum information because of the high speed and the weak decoherence during their transmission in channel. Usually quantum information is encoded in a two-dimensional space spanned for example by orthogonal polarizations of a photon or different paths along which a photon transmits. By this way, each photon could carry at most a quantum bit (qubit) information. If the photon could live in a high-dimensional space, for example, spanned by the inherently infinite-dimensional orbital angular momentums (OAM), then the information carried by each photon could be increased significantly, and the channel capacity of the network and the transmission efficiency could be also improved greatly [3, 4]. Moreover, in comparison to a two-dimensional state, high-dimensional states show many interesting properties: enable more efficient quantum-information processing, and afford a more secure flux of information in quantum key distribution [5], etc. Therefore quantum communication research based on a carrier lived in a high-dimensional space becomes a hot topic and attracts much attention recently, some quantum schemes using for example a photonic high-dimensional time-bin state [16] or a OAM state [17-19] have been reported.

Quantum repeaters are indispensable for increasing the transmission distance and improving the quantum information processing efficiency [20], among which a quantum memory is the key component consisting of the quantum repeater. If we could realize the reversible transfer of a high-dimensional quantum state between a true single photon and a matter used as a quantum memory with high fidelity and reliability, then we may have the potential solution in enhancing the channel capacity significantly in addition to overcoming distance limitations of quantum communication schemes through transmission losses, a high-dimensional quantum network may become practical. Therefore many groups and researchers are devoted to performing the storage of a light lived in a high-dimensional space. Although some works have reported on the storage of a light carrying OAM or a spatial structure in different physical systems, these works involve bright lights [21-28]. Very recently, Ref. 29 and 30 reported on the storage of a light at single-photon-level, imprinted an OAM state, but the photon is still in a two-dimensional space, carrying at most a qubit information. So far there is no any work reporting on the storage of a photon encoded to be a high-dimensional state in any physical system. Constructing such a quantum memory for a photonic qudit is still a big challenge.

Here, we report the first experimental realization of a quantum memory which could store a true single-photon encoded to be a high-dimensional quantum state. The photon stored is a heralded single photon prepared through the spontaneous four-wave mixing (SFWM) process via a double lambda configuration in a cold atomic ensemble. This photon carries a superposition state of OAM and is stored in another cold atomic ensemble via electromagnetically induced transparency (EIT). We reconstruct the storage process density matrix of a qutrit-quantum memory by the quantum process tomography with the fidelity of 85.3%, prove the feasibility of storing a three-dimensional state in this kind of the memory. Furthermore, as the examples, we perform the experimental storages of two special photonic qutrit states. Our results show that it is promising for realizing a high-dimensional quantum memory in the future.

**Results**

  **Reconstructing the density matrix of the quantum storage process.** The single photon used in the experiment was prepared using SFWM in a cold $^{85}$Rb atomic ensemble trapped in a two-dimensional magneto-optical trap (MOT) [31]. The experimental setup used in this work is the same as Ref. 30. The optical depth (OD) of MOT 1 measured was about 8. In the simplified experimental setup in Fig. 1(a), photon signal 1 at 780 nm is used as a trigger and photon signal 2 at 795 nm is stored for subsequent treatment; therefore, we hereafter call photon signal 1 the trigger and photon signal 2, the signal. The powers of pump 1 and pump 2 lasers are 660 μW and 45 μW respectively. We characterized the non-classical correlation between two photons in a pair and demonstrated the single-photon property of the signal photon using the method and procedure shown in Ref. 30. The experimental storage of the signal through EIT was implemented in another MOT 2. The OD of MOT 2 was about 20. The bandwidth for storage was about ~30 MHz. The Rabi frequency of the coupling laser was 4Γ, and the beam waist was of 2.5 mm. Compared with work in Ref. 30, we further improved the whole system, obtained an α (anti-correlation parameter) value of 0.006 for the signal before storage and of 0.08 for the retrieved signal from MOT 2. Both α values go to zero, confirming clearly that the single-photon nature is preserved during storage.

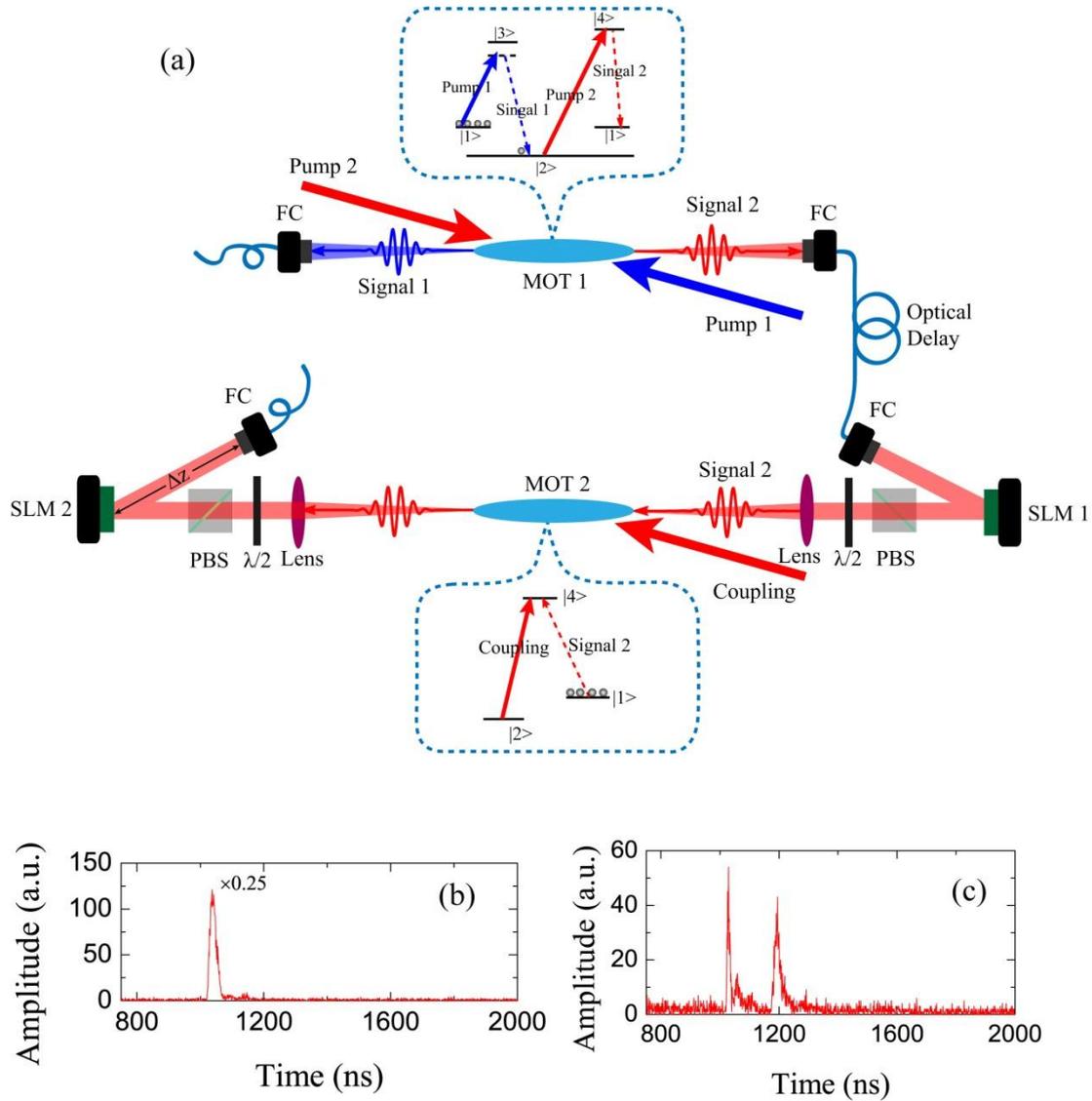

**Figure.1| Simplified experimental setup.** (a) Simplified diagram depicting the generation of non-classical photon correlations using SFWM and the storage of single photons. MOT: magneto-optical trap; Lens: lens with focus length of 300 mm; FC: fibre coupler; SLM: spatial light modulator; PBS: polarisation beam splitter; λ/2: half-wave plate. Inset: energy level diagrams for SFWM and EIT respectively; (b) Cross-correlation between the input signal and the trigger. (c) Cross-correlation function between the retrieved signal and the trigger after a programmed storage time.

We considered the quantum memory storing a three-dimensional state as an example to demonstrate the ability of storing a high-dimensional state in the cold atomic ensemble. The photonic qutrit lived in a space spanned by OAM states. Storing an input state and releasing it later can be considered as a state transfer process, and represented by a quantum process matrix $\chi$. For reconstructing the quantum storage process density matrix for a photonic qutrit, we performed the quantum process tomography: we input one of 9 different states $|\psi_{1,2,\sim 9}\rangle$ of $|L\rangle$, $|G\rangle$, $|R\rangle$, $(|G\rangle+|L\rangle)/2^{1/2}$, $(|G\rangle+|R\rangle)/2^{1/2}$, $(|G\rangle+i|L\rangle)/2^{1/2}$, $(|G\rangle-i|R\rangle)/2^{1/2}$, $(|L\rangle+|R\rangle)/2^{1/2}$, $(|L\rangle+i|R\rangle)/2^{1/2}$ to MOT 2 for storage respectively, where $|L\rangle$, $|G\rangle$ and $|R\rangle$ correspond to a well-defined OAM of 1 $\hbar$, 0 and -1 $\hbar$. These input states having different phase and intensity distributions as shown in Fig. 2(a), were prepared though a spatial light modulator (SLM 1) (HOLOEYE, PLUTO, with a resolution of 1920×1080. The SLM is reflective-type with a 60% efficiency.). After the programmed storage time, the stored state was retrieved and measured in nine different OAM basis vectors, the same with input 9 states [34]. These basis vectors for measurement were prepared by using another spatial light modulator (SLM 2), which were set to be in reversible phase rotations compared with the corresponding input state in order to obtain a Gaussian mode for measurement (see Method section).

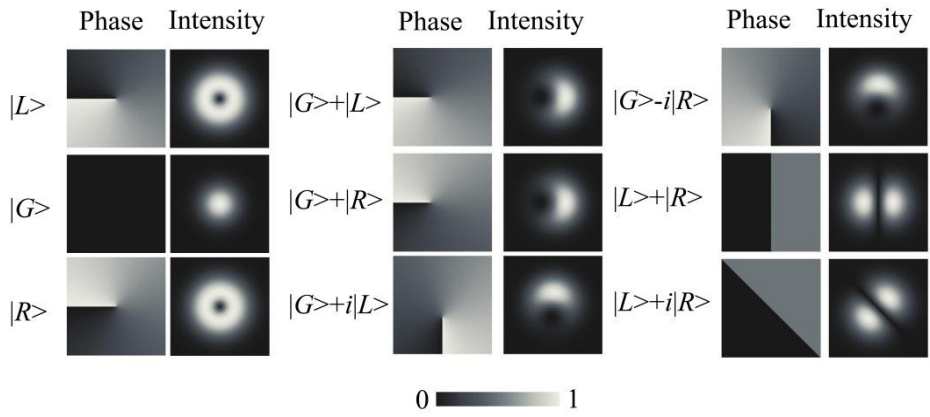

(a)

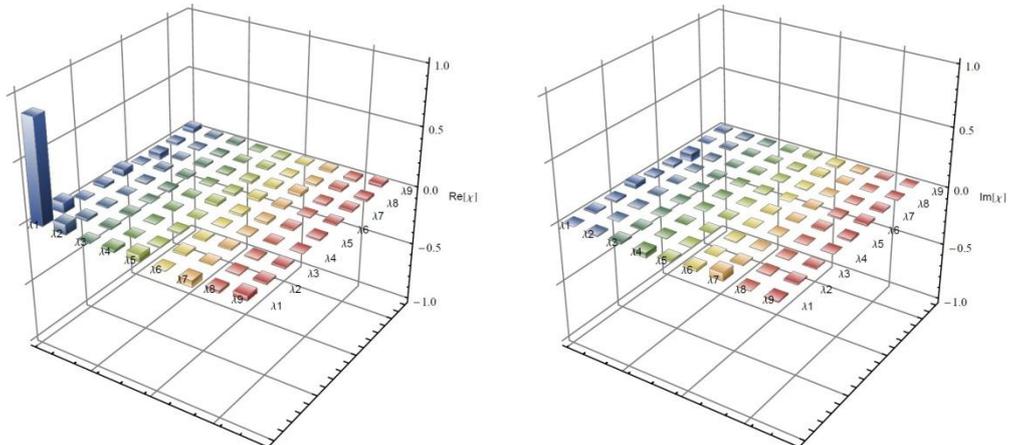

(b)                                                      (c)

**Figure.2| Reconstructing quantum storage process density matrix.** (a) The intensity (left panel of each pair) and the phase (right panel) distributions of input OAM states. (b) and (c) are the real and imaginary parts of reconstructed quantum storage process density matrix. $\lambda_{1-9}$ is a basis for operator, defined in Methods.

In addition, a 4-f imaging system was used in the experiment, which is the key for reconstructing correctly the density matrix: the SLM 1 acts as a mask plane, the centre of the atomic ensemble in MOT 2 is the Fourier plane and the SLM 2 represents the image plane, two lenses with the focus length of 300 mm are used to image SLM1 to SLM 2. Due to the conjugate properties between the mask plane and the image plane, the vectors for measurement need to be converted to be the conjugate to the corresponding input state (see Method). Another thing we want to point out is that the stored image in the atomic ensemble is the Fourier transformation of the input OAM state. The phase distributions at the image plane or the measurement plane are the Fourier transformations of the OAM states. So if we want to obtain the retrieved OAM state of the input photon, we have to Fourier transform the measured phase distribution of the retrieved photon, which is realized through a far-field diffraction by setting the single-mode fibre coupler used for collecting the retrieved photons 2.5 m away from the SLM 2 (Fig. 1(a)) in our experiment. In this process, the Fresnel diffraction could be ignored and the Fraunhofer diffraction is predominant. By inputting 9 input states respectively and measuring each retrieved input state in 9 basis vectors, and integrating coincidence counts in a 50-ns coincidence window with background noise subtracted in each measurement, we obtained a set of 81 data points and reconstructed the quantum process density matrix for our quantum memory system using them [35, 36]. The results were shown in Fig. 2 (b) and (c). Fig. 2(b) was the real part of the density matrix and Fig. 2(c) corresponded to the imaginary part. Compared with the ideal quantum storage process density matrix, the fidelity of our obtained density matrix was of 85.3%.

**Storing photonic qutrits.** Through reconstructing the quantum storage process density matrix, we concluded that our system can store a photonic qutrit. In order to further illustrate this ability of storage, we performed the experiments of storing two special photonic qutrits of $|\Psi_1\rangle=(|L\rangle+|G\rangle+|R\rangle)/3^{1/2}$ and $|\Psi_2\rangle=(|L\rangle-|G\rangle+|R\rangle)/3^{1/2}$ as examples. The phase structures and the intensity distributions of these two states are given in Figs. 3(a) and 3 (d). Through projecting these two states on nine basis vectors defined before, we obtained the nine coincidence counts respectively and reconstructed the density matrix of the retrieved state using them, shown in Figs. 3(b-c) and 3(e-f). Fig. 3(b) and Fig. 3(e) were the real parts of retrieved photonic qutrit states. Fig. 3(c) and Fig. 3(f) corresponded to the imaginary parts respectively. We calculated the fidelity of the reconstructed density matrix by comparing with the ideal density matrix, which are 77%±3% for state $|\Psi_1\rangle$ and 80%±2% for state $|\Psi_2\rangle$ respectively.

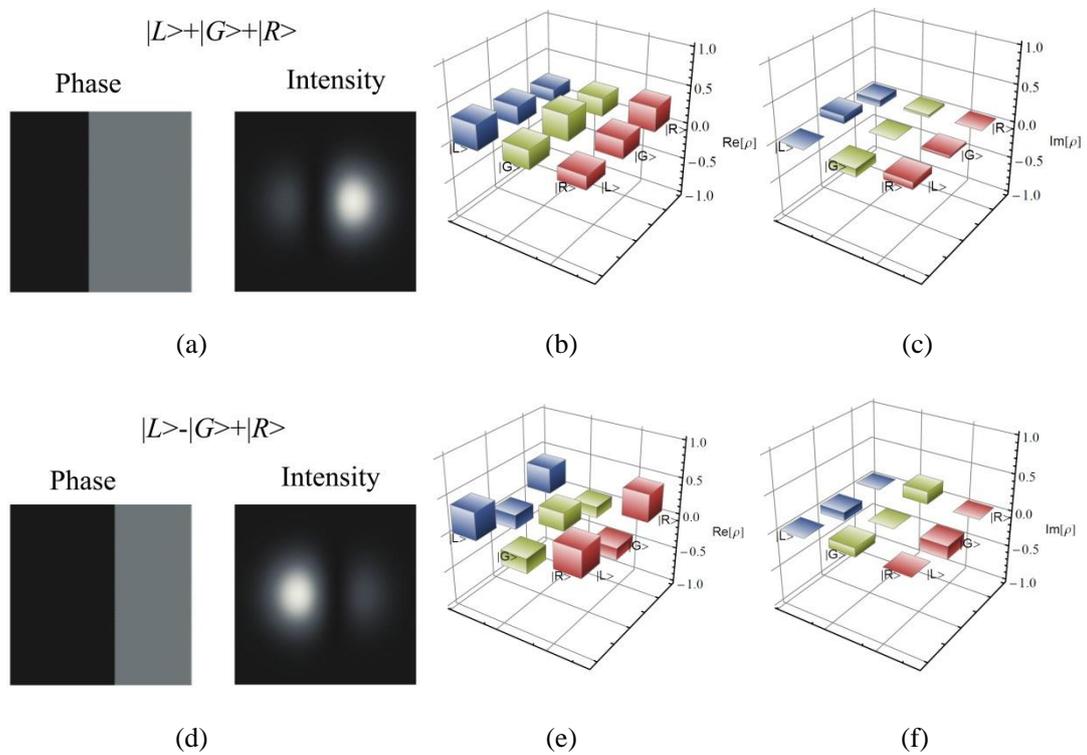

**Figure.3| Storing photonic qutrits.** (a) The phase (left panel of each pair) and the intensity (right panel) distributions of the photonic qutrit state $|\Psi_1\rangle=(|L\rangle-|G\rangle+|R\rangle)/3^{1/2}$. (b) and (c) are the real and imaginary parts of the reconstructed density matrix of the retrieved photonic qutrit $|\Psi_1\rangle$. (d) The phase (left panel of each pair) and the intensity (right panel) distribution of the photonic qutrit state $|\Psi_2\rangle=(|L\rangle-|G\rangle+|R\rangle)/3^{1/2}$. (e) and (f) are the real and imaginary parts of reconstructed density matrix of retrieved photonic qutrit state $|\Psi_2\rangle$.

**Discussions**

Although our work is a proof of principle for storing a photonic qutrit, it is the first step towards realizing a high-dimensional quantum memory. Moreover, our method could be extended to reconstruct quantum process density matrix of storing a high-dimensional OAM state ($d \geq 4$) according to Ref. [35]. In this case, the big challenges we face in the experimental demonstrations are further techniques improvements, such as how to achieve higher signal-to-noise ratio, to make the system keep stable over long time, etc. This is because we have to take more data in the experimental demonstration. For example, for $d=4$, 256 data points should be obtained for reconstructing the process density matrix, which needs longer experimental time.

The fidelity of the storage process is lower than 100%, we think the main reasons are follows: 1. in the experiment, we use a 4-f imaging for measuring the retrieved state. The mismatch between the position signal beam exposures at the SLM 2 in fact and the ideal position the signal should exposure reduces the fidelity. One could soften this influence by enlarging the beam waist of the signal. 2. When the input state is $|\psi_8\rangle$ or $|\psi_9\rangle$ state or the projecting of any input state in basis vector $|\psi_8\rangle$ or $|\psi_9\rangle$, we have to seek the help of the far-field diffraction by setting the single-mode fibre coupler used for collecting the retrieved photons far away from the SLM 2. The relative small distance of 2.5 m may not completely project the input state into the Gaussian mode, which also induces the errors and reduces the fidelity. The same reasons affect on the fidelity of the photonic qutrit states used as examples more seriously, because a three-dimensional state prepared by SLM is used as an input in these cases, compared with the case of reconstructing the process density matrix, where a two-dimensional state is used.

**Method Section**

**Performing quantum process tomography.** In discrete-variable quantum information, storing a state and retrieving it later can be regarded as a state transfer process. This process can be represented by a quantum process matrix $\chi$. The output state $\varepsilon(\rho)$ can be written as:

$$\varepsilon(\rho) = \sum_{m,n=1}^{9} \chi_{mn} \hat{\lambda}_m \rho \hat{\lambda}_m^\dagger \tag{1}$$

Where $\hat{\lambda}_m$ is a basis for operator acting on input state $\rho$. The matrix $\chi$ can be obtained by measuring the output state $\varepsilon(\rho)$. The complete operators for reconstructing matrix $\chi$ of a photonic qutrit are given: 
$$\lambda_1 = \begin{pmatrix} 1 & 0 & 0 \\ 0 & 1 & 0 \\ 0 & 0 & 1 \end{pmatrix}, \quad \lambda_2 = \begin{pmatrix} 0 & 1 & 0 \\ 1 & 0 & 0 \\ 0 & 0 & 0 \end{pmatrix}, \quad \lambda_3 = \begin{pmatrix} 0 & -i & 0 \\ -i & 0 & 0 \\ 0 & 0 & 0 \end{pmatrix}, \quad \lambda_4 = \begin{pmatrix} 1 & 0 & 0 \\ 0 & -1 & 0 \\ 0 & 0 & 0 \end{pmatrix},$$

$$\lambda_5 = \begin{pmatrix} 0 & 0 & 1 \\ 0 & 0 & 0 \\ 1 & 0 & 0 \end{pmatrix}, \quad \lambda_6 = \begin{pmatrix} 0 & 0 & -i \\ 0 & 0 & 0 \\ i & 0 & 0 \end{pmatrix}, \quad \lambda_7 = \begin{pmatrix} 0 & 0 & 0 \\ 0 & 0 & 1 \\ 0 & 1 & 0 \end{pmatrix}, \quad \lambda_8 = \begin{pmatrix} 0 & 0 & 0 \\ 0 & 0 & -i \\ 0 & i & 0 \end{pmatrix}, \quad \lambda_9 = \begin{pmatrix} 1 & 0 & 0 \\ 0 & 1 & 0 \\ 0 & 0 & -2 \end{pmatrix}/\sqrt{3}.$$

The input states are: $|\psi_1\rangle = \begin{pmatrix} 1 \\ 0 \\ 0 \end{pmatrix}$, $|\psi_2\rangle = \begin{pmatrix} 0 \\ 1 \\ 0 \end{pmatrix}$, $|\psi_3\rangle = \begin{pmatrix} 0 \\ 0 \\ 1 \end{pmatrix}$, $|\psi_4\rangle = \begin{pmatrix} 1 \\ 1 \\ 0 \end{pmatrix}/\sqrt{2}$, $|\psi_5\rangle = \begin{pmatrix} 0 \\ 1 \\ 1 \end{pmatrix}/\sqrt{2}$,

$|\psi_6\rangle = \begin{pmatrix} i \\ 1 \\ 0 \end{pmatrix}/\sqrt{2}$, $|\psi_7\rangle = \begin{pmatrix} 0 \\ 1 \\ i \end{pmatrix}/\sqrt{2}$, $|\psi_8\rangle = \begin{pmatrix} 1 \\ 0 \\ 1 \end{pmatrix}/\sqrt{2}$, $|\psi_9\rangle = \begin{pmatrix} 1 \\ 0 \\ i \end{pmatrix}/\sqrt{2}$. In our experiment, these states corresponds OAM states of |L>, |G>, |R>, (|G>+|L>)/$2^{1/2}$, (|G>+|R>)/$2^{1/2}$, (|G>+i|L>)/$2^{1/2}$, (|G>-i|R>)/$2^{1/2}$, (|L>+|R>)/$2^{1/2}$, (|L>+i|R>)/$2^{1/2}$ respectively, where |L>, |G> and |R> are the single photon states with OAM of $-\hbar$, $0$, $+\hbar$ respectively. By storing these nine input states in MOT 2 respectively and measuring the corresponding retrieved output state in 9 basis vectors represented by the operators $\hat{\mu}_i \otimes \hat{\mu}_j$ ($i, j$=1,2,···9, $\hat{\mu}_i = |\psi_i\rangle\langle\psi_i|$), we can reconstruct the quantum storage process matrix $\chi$ according to Eq. (1). The fidelity of this quantum process can be calculated by an equation of $F_{\text{Process}} = Tr(\sqrt{\sqrt{\chi}\chi_{\text{ideal}}\sqrt{\chi}})^2$, where $\chi_{\text{ideal}}$ is the ideal quantum storage process density matrix.

**Imaging the input spatial state.** In our experiment, SLM 1, MOT 2 and SLM 2 and two lenses consist of a 4-f imaging system: SLM 1 as a mask plane, the atomic ensemble in MOT 2 is the Fourier plane and the SLM 2 represents the image plane. Due to the conjugate properties between the mask plane and the image plane, the selected measurement vectors need to be converted to be the conjugate of the corresponding input states. In order to illustrate the imaging process, we give the phase distributions in mask, Fourier and image planes in Fig. 4. The columns (1) show the different phase distributions imaged on the SLM 1, the columns (2) represents the phase distributions in Fourier plane by imaging the phase distribution on SLM 1. The columns (3) are the selected eigenvectors for measurement programmed on SLM 2. These phase structures are modulated by two spatial light modulators.

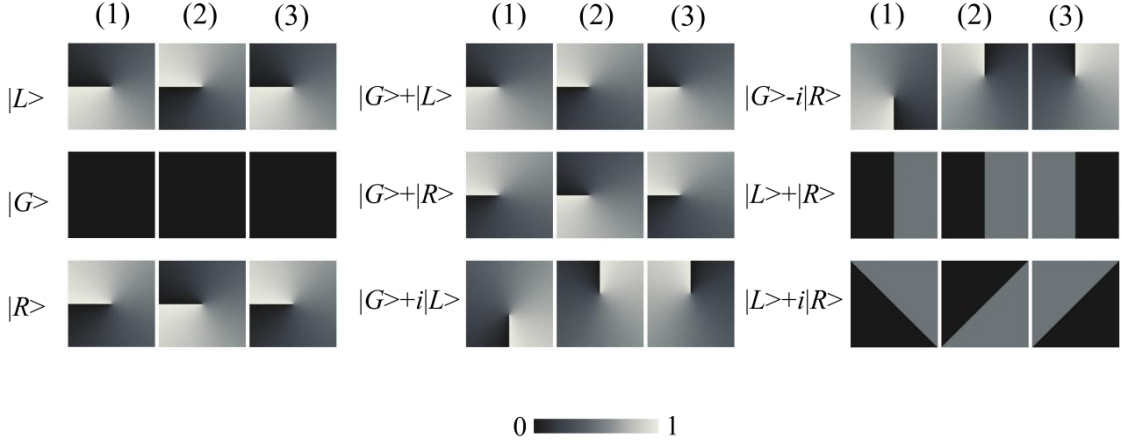

**Fig. 4 The phase distributions in mask, Fourier and image planes.** The column (1) shows the phase distributions imaged on the SLM 1; the column (2) represents the phase distributions in Fourier plane; the column (3) is the selected eigenvector for measurement programmed on SLM 2

**Acknowledgements**

We thank Dr. Xian-Min Jin, Dr. Xi-Feng Ren and Dr. Jin-Shi Xu for helpful discussions. This work was supported by the National Natural Science Foundation of China (Grant Nos. 11174271, 61275115, 10874171), the National Fundamental Research Program of China (Grant No. 2011CB00200), the Youth Innovation Fund from USTC (Grant No. ZC 9850320804), and the Innovation Fund from CAS, the program for NCET.

**Author contributions**

BSS and DSD conceived the experiment for discussion. The experimental work and data analysis were carried out by DSD and BSS, with assistance from WZ, ZYZ. The theoretical analysis is performed by DSD and BSS with assistance from JSP and GYX. BSS and DSD wrote this paper. BSS and GCG supervised the project.